**ABSTRACT WORD COUNT:** 299

**MANUSCRIPT WORD COUNT**: 3,416

**TITLE: Gamified Crowdsourcing as a Novel Approach to Lung Ultrasound Dataset Labeling**

**RUNNING TITLE: Crowdsourcing Lung Ultrasound Labeling**

**MANUSCRIPT TYPE:** Original Investigation

**AUTHORS:**

1. Nicole M Duggan*, MD (nmduggan@bwh.harvard.edu)[a]

2. Mike Jin*, PhD (mike@centaurlabs.com)[bc]

3. Maria Alejandra Duran Mendicuti, MD (mduranmendicuti@bwh.harvard.edu)[c]

4. Stephen Hallisey, MD (shallisey@partners.org) [a]

5. Denie Bernier, RDMS (dbernier1@bwh.harvard.edu)[a]

6. Lauren A Selame, MD (lselame@bwh.harvard.edu)[a]

7. Ameneh Asgari-Targhi, PhD (aasgaritarghi@bwh.harvard.edu)[c]

8. Chanel E Fischetti, MD (cfischetti@bwh.harvard.edu)[a]

9. Ruben Lucassen, MS (r.t.lucassen@tue.nl)[d]

10. Anthony E Samir, MD, MPH (asamir@mgh.harvard.edu)[e]

11. Erik Duhaime+, PhD (erik@centaurlabs.com)[b]

12. Tina Kapur+, PhD (tkapur@bwh.harvard.edu)[c]

13. Andrew J Goldsmith+, MD, MBA (ajgoldsmith@bwh.harvard.edu)[a]

*These authors contributed equally to primary authorship.

+These authors contributed equally to senior authorship.

**AUTHOR AFFILIATIONS:**

[a] Department of Emergency Medicine, Brigham and Women's Hospital, Harvard Medical School, Boston, MA, USA

[b] Centaur Labs, Boston, MA, USA

[c] Department of Radiology, Brigham and Women's Hospital, Harvard Medical School, Boston, MA, USA

[d] Department of Biomedical Engineering, Eindhoven University of Technology, The Netherlands

[e] Department of Radiology, Massachusetts General Hospital, Harvard Medical School, Boston, MA, USA

**CORRESPONDING AUTHOR:**

Nicole M. Duggan, MD

75 Francis Street, NH-2

Boston, MA 02115, USA

Email: nmduggan@bwh.harvard.edu

**CONFLICT OF INTEREST DISCLOSURES**

NMD reports no disclosures.

MJ is a current salaried employee and holds equity in Centaur Labs.

MADM reports no disclosures.

SH reports no disclosures.

DB reports no disclosures.

LAS reports no disclosures.

AAT reports no disclosures.

CEF is a former salaried employee and currently holds equity in Centaur Labs. She has served as a compensated consultant for Philips and Level Ex.

RL reports no disclosures.

AES has served as a compensated consultant for Astra Zeneca, Bracco Diagnostics, Bristol Myers Squibb, General Electric, Gerson Lehman Group, Guidepoint Global Advisors, Supersonic Imagine, Novartis, Pfizer, Philips, Parexel Informatics, and WorldCare Clinical. He holds stock options in Rhino Healthtech Inc. and Ochre Bio Inc. He holds advisory board or committee memberships for General Electric, the Foundation for the National Institutes of Health, and the Sano Center for Computational Personalized Medicine, and is a member of the Board of Governors of the American Institute for Ultrasound in Medicine. His institution has received research grant support and/or equipment for projects that he has led from the Analogic Corporation, Canon, Echosens, General Electric, Hitachi, Philips, Siemens, Supersonic Imagine/Hologic, Toshiba Medical Systems, the US Department of Defense, Fujifilm Healthcare, the Foundation for the National Institutes of Health, Partners Healthcare, Toshiba Medical Systems, and Siemens Medical Systems. He holds equity in Avira Inc., Autonomus Medical Technologies, Inc., Evidence Based Psychology LLC, Klea LLC, Katharos Laboratories LLC, Quantix Bio LLC, and Sonoluminous LLC. He receives royalties from Elsevier Inc. and Katharos Laboratories LLC.

ED is the Chief Executive Officer of and holds equity in Centaur Labs.

TK reports no disclosures.

AJG has served as a compensated consultant for Philips and Exo.

**FUNDING INFORMATION**

Massachusetts Life Sciences Center

**ACKNOWLEDGEMENTS:**

Drs. Kapur and Goldsmith had full access to all of the data in the study and take responsibility for the integrity of the data and the accuracy of the data analysis.

*Concept and design*: Duggan, Jin, Duhaime, Kapur, Goldsmith

*Acquisition, analysis, or interpretation of the data*: Duggan, Jin, Duran Mendicuti, Hallisey, Bernier, Selame, Duhaime, Kapur, Goldsmith

*Drafting of the manuscript*: Duggan, Jin, Duhaime, Kapur, Goldsmith

*Critical revision of the manuscript for important intellectual content*: Duggan, Jin, Duran Mendicuti, Hallisey, Bernier, Selame, Asgari-Targhi, Fischetti, Lucassen, Samir, Duhaime, Kapur, Goldsmith

*Statistical analysis*: Jin, Lucassen, Kapur,

*Obtained funding*: Duhaime, Kapur, Goldsmith, Duggan

*Administrative, technical, or material support*: Fischetti, Samir, Lucassen

*Supervision*: Duhaime, Kapur, Goldsmith

**ABSTRACT**

**Study Objective:** Machine learning models have advanced medical image processing and can yield faster, more accurate diagnoses. Despite a wealth of available medical imaging data, high-quality labeled data for model training is lacking. We investigated whether a gamified crowdsourcing platform enhanced with inbuilt quality control metrics can produce lung ultrasound clip labels comparable to those from clinical experts.

**Methods:** 2,384 lung ultrasound clips were retrospectively collected from 203 patients. Six lung ultrasound experts classified 393 of these clips as having no B-lines, one or more discrete B-lines, or confluent B-lines to create two sets of reference standard labels (195 training set clips and 198 test set clips). Sets were respectively used to A) train users on a gamified crowdsourcing platform, and B) compare concordance of the resulting crowd labels to the concordance of individual experts to reference standards.

**Results:** 99,238 crowdsourced opinions on 2,384 lung ultrasound clips were collected from 426 unique users over 8 days. On the 198 test set clips, mean labeling concordance of individual experts relative to the reference standard was 85.0% ± 2.0 (SEM), compared to 87.9% crowdsourced label concordance (p=0.15). When individual experts' opinions were compared to reference standard labels created by majority vote excluding their own opinion, crowd concordance was higher than the mean concordance of individual experts to reference standards (87.4% vs. 80.8% ± 1.6; p<0.001).

**Conclusion:** Crowdsourced labels for B-line classification via a gamified approach achieved expert-level quality. Scalable, high-quality labeling approaches may facilitate training dataset creation for machine learning model development.

**INTRODUCTION**

Machine learning (ML) models can improve medical diagnostic accuracy and streamline healthcare processes.[1] This is particularly true when applied to medical image analysis. ML models require large-scale labeled datasets for model training.[2] Widespread ML tool development is limited by the need to acquire high-quality labels given associated expert time and cost.[3-7]

Combining opinions from multiple individuals on a given task can produce more accurate interpretations than from a single individual.[8] Crowdsourcing, the process of collecting large numbers of unskilled opinions, can improve efficiency, lower costs, and offer high-quality in repetitive task completion.[8-9] Crowdsourced approaches to dataset labeling are growing in popularity, and beneficial effects of crowdsourcing have been demonstrated in healthcare-related tasks including biomedical imaging analysis.[10-14]

Using crowdsourcing for biomedical image labeling is challenged by the complexity of the tasks and the need to ensure label quality control. The user interface design for collecting crowd opinions and the metrics used for assessing opinion quality are key to successful results. Gamification, the persuasive system design which uses game-like tasks to engage participants competitively for rewards, can both encourage crowd participation and improve performance accuracy by selectively rewarding top users.[15-18] Combining crowdsourcing with gamification can become a performance measurement tool for identifying top crowd users.[18-21]

Point-of-care ultrasound (POCUS) is a dynamic medical imaging technique used at patients' bedside to make accurate, real-time diagnoses.[22-24] Though POCUS has significant value in healthcare settings, advanced training is required to accurately apply this tool to clinical care.[25] As such, ML models which automate POCUS image interpretation hold exceptional potential clinical value. In lung POCUS, B-lines are hyperechoic linear artifacts that extend from the pleural line and appear dynamically with the respiratory cycle.[23] B-lines are known markers of pulmonary congestion and their presence, quantity, and thickness (discrete vs confluent B-lines) correlate with pathological severity of conditions such as congestive heart failure exacerbations, pneumonia, or inflammatory lung disease.[23,26,27] Early ML models for lung POCUS have shown promise in identifying and quantifying the presence of B-lines.[28-33] However existing models have limited accuracy and generalizability due in part to the lack of large-scale, high-quality, labeled image databases for model training. Gamified crowdsourcing has not previously been applied to annotate lung POCUS clips or ultrasound imaging data overall.

**Goals of This Investigation**

We examined whether a gamified crowdsourcing approach with inbuilt quality control measures can classify lung POCUS clips for presence and type of B-lines at comparable accuracy to trained ultrasound experts. Our secondary objectives were to measure (a) the number of crowd opinions required to achieve maximal accuracy, and (b) learning curves for expert and crowd individuals over time.

**MATERIALS AND METHODS**

**Study Design and Setting**

This was a prospective analysis performed using retrospectively collected lung POCUS clips. All lung POCUS examinations performed in an academic tertiary care emergency department between March 1st, 2020 and February 28th, 2022 were retrospectively queried via the electronic health record. This study was approved by the local institutional review board.

**Dataset Curation**

In total, 2,391 POCUS clips were downloaded in Digital Imaging and Communications in Medicine (DICOM) format from the hospital Picture Archiving and Communication System. Six-second clips were acquired at frame rates between 15-46 Hz. DICOM files were converted to MP4 format using an open-source medical image viewer and subsequently de-identified using a software package.[34,35]

The 2,391 clips were randomly divided by patient into two sets: dataset A (102 patients, 1,271 clips) and dataset B (101 patients, 1,120 clips). 200 random clips from dataset A were selected as a crowd training set, and 200 random clips from dataset B were selected as a test set to evaluate crowd label quality. 5 training set clips and 2 test set clips were excluded for being flagged by at least one expert as not containing lung (Figure 1A).

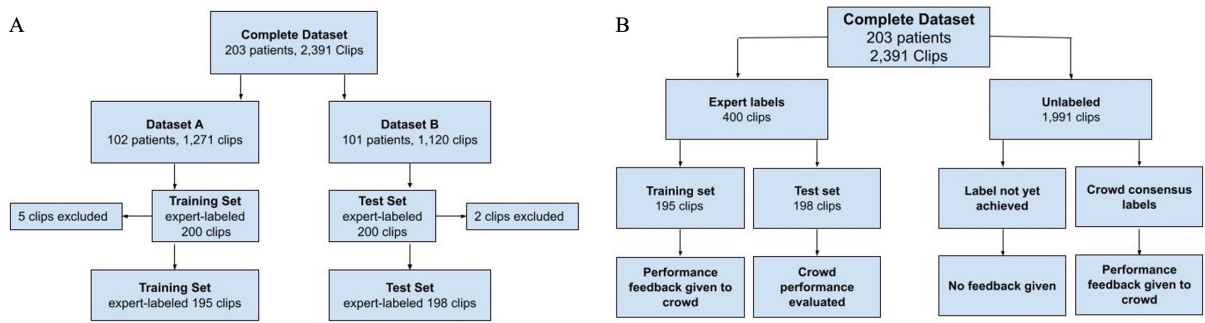

**Figure 1.** Dataset processing flow diagram. **(A)** Dataset partitioning. The complete dataset was divided evenly by patient into Dataset A and Dataset B. 200 clips from each dataset were randomly selected to create a training set and test set, respectively. Five clips from the training set and two clips from the test set were excluded for having no lung visible in the clip prior to deploying these sets to the crowd. **(B)** Lung POCUS clip lifecycle. Of the complete dataset, 400 clips were separated out to become the training set, used for giving performance feedback to the crowd, and the test set, use to evaluate crowd performance and accuracy. The remaining clips from the complete dataset were not labeled by experts, but instead were available to achieve a crowd-consensus label. Crowd users were given additional performance feedback on clips that had previously achieved a crowd-consensus label.

**Task Definition and Reference Standards**

Expert and crowd users were asked to classify B-lines on lung POCUS clips into one of three classes: a) no B-lines, b) one or more discrete B-lines, or c) confluent B-lines (Figure 2).

Clips were classified based on the highest B-line severity present throughout the entire clip, with no B-lines < one or more discrete B-lines < confluent B-lines. Discrete B-lines were defined as hyperechoic lines originating from the pleural line, demonstrated sliding with the pleura, and extending to the bottom of the sonographic field.[36,37] Confluent B-lines were defined as hyperechoic sections originating from the pleural line, demonstrated sliding with the pleura, and had thickness along the pleura beyond that of discrete B-lines.[36,37]

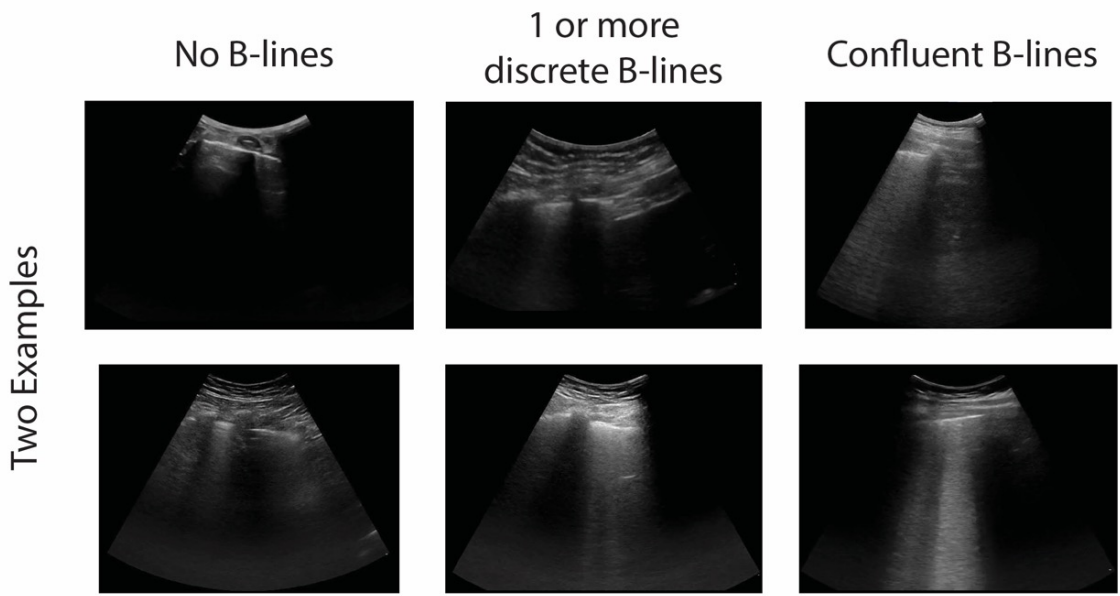

**Figure 2.** Representative frames from three LUS images demonstrating the three classification categories. **(A)** No B-lines; **(B)** Discrete B-lines; **(C)** Confluent B-lines.

Six experts with advanced training in lung POCUS (four ultrasound fellowship-traiined emergency medicine physicians including one dual board-certified in intensive care; one emergency radiologist, one Registered Diagnostic Medical Sonographer) were recruited. All six experts provided independent classification opinions for all training and test set clips via DiagnosUs (Centaur Labs, Boston, MA), a free iOS application where users compete in medical data labeling contests to win cash prizes based on their labeling accuracy. All clips could be continuously played or viewed frame-by-frame as many times as needed. Prior to participation, experts underwent a 30-minute training session familiarizing themselves with the application and ensuring agreement on the study-specific classifications.

Reference standard labels on training and test set clips were assigned using expert consensus – the majority rule of the six experts' opinions, with ties broken randomly. Reference standard labels using only five experts' opinions at a time were also created for "leave-one-out" label concordance analysis (see Appendix).

**Gamified Crowdsourcing**

Crowd opinions were collected using DiagnosUs via gamified contests. Crowd users included anyone in the general public with access to an iOS device who downloaded the iOS application. There were no criteria regarding level of expertise to participate. Users voluntarily participated in labeling contests based on interest and the potential to earn rewards.

Crowd users were trained in two ways: A) optional tutorial cases with accompanying layperson explanations which they could access at any time, and B) after submitting an opinion, immediate feedback was given via revelation of the current label (i.e., the reference standard label on training set clips, or the current crowd consensus label if one exists). Crowd consensus labels were computed and assigned for all clips using the majority rule of the top crowd labelers' opinions, and were assigned to any of the 1,991 initially unlabeled clips once sufficient mutual agreement was reached amongst the top crowd labelers for that clip (Figure 1B). Top labelers were identified via continuous performance monitoring based on their trailing accuracy on clips with labels. Clips were shown to users in random order so that they could not predict which clips would impact their contest score, and any test clip could be shown to a user multiple times. For additional details on crowdsourcing mechanics (see Appendix).

**Crowd Consensus Label Quality Analysis**

Crowd label quality was assessed by comparing how well crowd consensus labels and individual expert opinions respectively matched the reference standard labels on the test set. Both of these were measured by concordance, calculated simply as the percent of matching labels across the test set clips. In addition, we also calculated concordances with "leave-one-out" reference standards. A collection of six "leave-one-out" reference standard label sets was created based on five experts each, one for each expert whose input was excluded from the majority rule consensus. This was done in order to measure each expert against a reference standard which their own opinion did not influence. See Appendix for additional details.

**Analysis of Secondary Outcomes**

To assess the minimum number of crowd opinions needed to achieve expert-level accuracy, we estimated the accuracy that the crowd labels would have had if we had collected fewer opinions on each clip by repeatedly sampling fewer opinions from all test clips and computing concordance with the reference standard based on the resulting crowd consensus labels. For every opinion count we used 1000 Monte Carlo samples per clip to estimate accuracy.

To calculate learning curves, the concordance of individual crowd users or experts with the reference standard was computed each time the individual gave an opinion on a test set clip, based on the proportion of their last 25 opinions on test set clips that matched the reference standard. The concordance at each opinion submission was averaged across individuals (all crowd users, skilled crowd users, or all experts) to produce the learning curves. Skilled crowd users were users who submitted at least one opinion on at least one test clip when they had a trailing average concordance (with respect to reference standards for training clips and crowd consensus labels for initially unlabeled clips) of 80% or better. For this analysis all opinions from crowd users were considered, including multiple opinions per user per clip if applicable.

**Statistical Analysis**

Analysis was performed using Python 3.10.[38] The difference between crowd concordance and average individual expert concordance was tested for significance using a paired samples *t*-test. All mean calculations are reported as mean ± standard error of mean (SE).

**RESULTS**

**Dataset Characteristics**

The patients who contributed to the lung POCUS database had a mean age of 60.0 years (standard deviation 19.0). 105 (51.7%) were female; 43 (21.2%) were Hispanic, 42 (20.7%) were Black, and 114 (56.1%) were White. From the ED, 64% of patients were admitted to the hospital floor, 28.6% were discharged home, 6.4% were admitted to the intensive care unit, 1% went directly to the operating room, 4.4% had an alternate disposition and 0% expired in the ED (Table 1).

Table 1. Patient characteristics for the collected dataset.

| Characteristic | Subjects, N = 203 [no. (%)] |
|---|---|
| **Sex** | |
| Male | 98 (48.3) |
| Female | 105 (51.7) |
| **Ethnicity** | |
| Hispanic or Latino | 43 (21.2) |
| Not Hispanic or Latino | 160 (78.8) |
| **Race** | |
| American Indian or Alaskan Native | 0 (0) |
| Asian | 9 (4.4) |
| Native Hawaiian or Pacific Islander | 0 (0) |
| Black or African American | 42 (20.7) |
| Caucasian | 114 (56.2) |
| Other | 38 (18.7) |
| **Emergency Department Disposition** | |
| Discharge home | 57 (28.6) |
| Floor admission | 122 (64.0) |
| Intensive care unit admission | 13 (6.4) |
| Operating room | 2 (1.0) |
| Other | 9 (4.4) |

**Opinion Collection**

Experts spent an average of 1.7 hours (minimum 0.9 hours, maximum 2.5 hours) submitting opinions for all 393 training and test clips (3.9 opinions per minute on average). Over the 195 training clips, the reference standard label distribution (based on the experts' majority opinion) was 58% no B-lines, 29% discrete B-lines, and 13% confluent B-lines. Over the 198 test clips the reference standard label distribution was 70% no B-lines, 18% discrete B-lines, and 12% confluent B-lines.

In total, 99,238 crowdsourced opinions were collected from 426 unique users across all 2,384 clips. The number of users contributing an opinion to each test set clip ranged from 28 to 48. Of these, 34,363 opinions from 114 unique users were eligible to be considered for crowd consensus labels based on our quality thresholds (see Crowd User Quality Assessment in Appendix). Of all users, 45% reported having prior medical experience compared to 53% of users who contributed to crowd consensus labels (p=0.061). The live contest was launched over 138 hours which reflected a mean acquisition rate of 12.0 opinions per minute. The total cash prize payout throughout the entire competition was 1,100 USD. The maximum prize earned by an individual user was 25 USD.

**Label Concordance With Reference Standard**

The six experts' concordances on the 198 test clips relative to the reference standard were 77.2%, 81.3%, 84.8%, 87.3%, 88.4%, and 90.9%, with a mean of 85.0 ± 2.0. Comparatively, the crowd concordance on these clips was 87.9% relative to the reference standard (p=0.15) (Figure 3A). When individual expert concordances were computed against "leave-one-out" reference standards, expert concordances on the same 198 test clips were 75.8%, 77.8%, 79.8%, 81.8%, 83.3%, and 86.4% with a mean of 80.8 ± 1.6, compared to a crowd concordance of 87.4% (p<0.001) (Figure 3B).

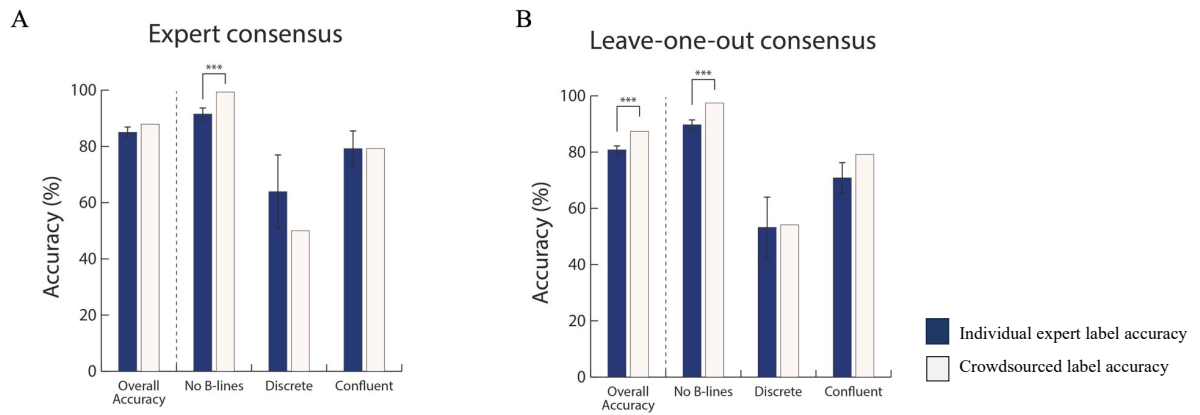

**Figure 3.** Expert and crowd accuracy stratified by B-line classifications. **(A)** Compared to the consensus of 6 experts. **(B)** Compared to the consensus of 5 experts (i.e., leave-one-out). ***: p<0.001.

For clips designated by the reference standard as having no B-lines, experts had an average concordance with the reference standard of 91.5% ± 2.3, compared to a crowd concordance of 99.3% (p<0.001). For cases with discrete B-lines, experts had an average concordance of 63.9% ± 13.2 compared to the crowd concordance of 50% (p=0.088). For cases with confluent B-lines, expert average concordance (79.2% ± 6.5) and crowd concordance were both 79.2% (p=1.0). The balanced multiclass accuracy (i.e., average per-class concordance) was 76.1% for the crowd and 78.2% ± 4.0 for individual experts relative to the reference standard, and 76.9% for the crowd and 71.2% ± 3.2 for individual experts relative to leave-one-out reference standard.

Using randomly sampled subsets of collected opinions, 7 quality-filtered opinions were sufficient to achieve near the maximum crowd accuracy (Figure 4).

**Expert and Crowd Labeling Disagreements**

Expert opinions were unanimous on 50.2% of training cases and 52.5% of test cases. There was an expert supermajority (i.e., two-thirds) on 89.2% of training cases and 87.9% of test cases. On average, 41.2 (σ=4.1) crowd opinions contributed toward the crowd consensus label for each of the test cases. Crowd opinions were unanimous on 35.4% of cases, and a crowd supermajority existed for 85.9% of test cases.

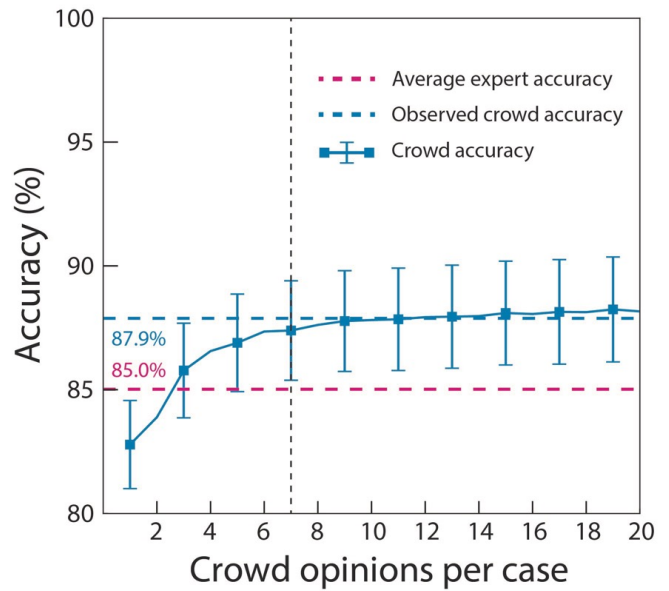

**Figure 4.** The number of crowd opinions needed to maximize crowd-consensus accuracy. Solid blue line indicates estimated crowd-consensus accuracy as dependent on the number of crowd opinions collected. Vertical dotted line indicates 7 crowd opinions are sufficient to achieve the observed crowd accuracy. Observed crowd accuracy indicates the overall crowd-consensus accuracy from using all crowd opinions on each clip. Error bars indicate standard error of the mean.

The receiver operating characteristic curves based on using different thresholds of crowd opinion proportions to predict the expert-consensus label produced high area under the curve (AUC) values for all B-line classifications: the AUC was 0.98 for the "No B-lines" classification, 0.95 for "discrete B-lines", and 0.98 for "confluent B-lines" (Figure 5).

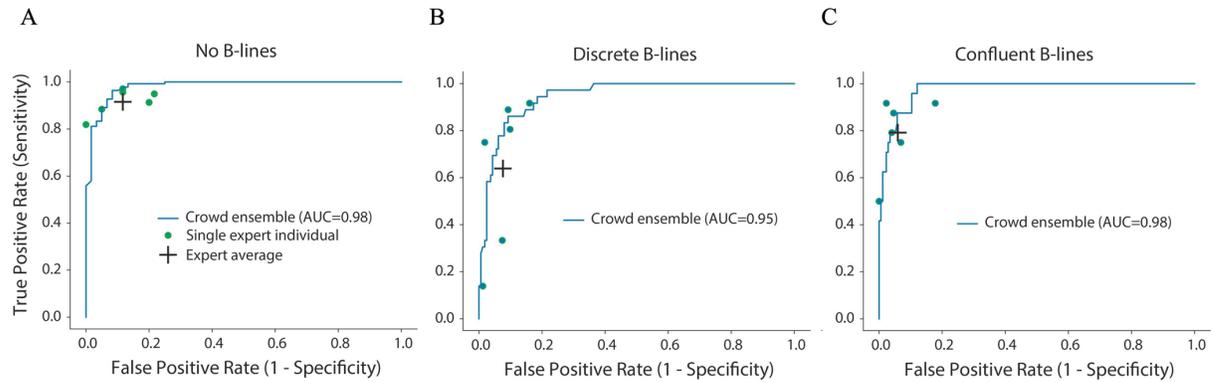

**Figure 5.** ROC curves by B-line classification for crowd opinions relative to the expert-consensus label. Sensitivity and specificity for individual experts are plotted as green points, and the average sensitivity and specificity of all six experts is shown as a black cross. **(A)** No B-lines; **(B)** Discrete B-lines; **(C)** Confluent B-lines.

Clips with discrete B-lines had the most disagreement from both the crowd consensus and individual experts with the expert consensus, making up 75% of cases where the crowd consensus differed from the expert consensus reference standard (Figure 6).

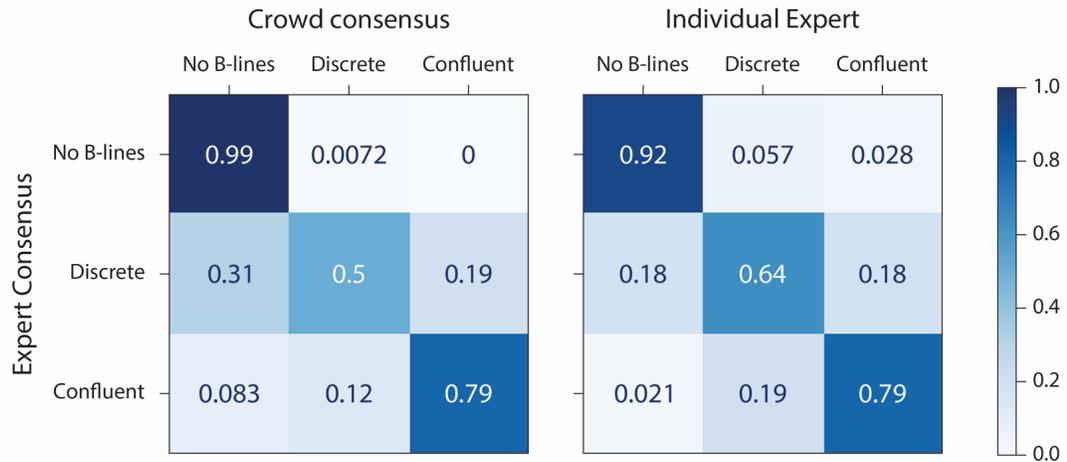

**Figure 6.** Confusion matrices showing disagreement by category between the expert-consensus and **(A)** the crowd-consensus, and **(B)** individual experts.

On a clip-by-clip basis, the level of internal crowd agreement and internal expert agreement on each test clip was significantly correlated (Pearson's r=0.70, p<0.001). When considering only test clips where the crowd opinions were at least 80% in agreement, the crowd consensus label concordance with the reference standard was 96%. Expert agreement was significantly higher on test clips where the crowd label matched the reference standard (89%) versus on test clips where it did not (60%) (p<0.001, Mann-Whitney *U* test).

**Expert and Crowd Learning**

Individual crowd users showed improved concordance with the reference standard over time as each user gave opinions on more clips and thus received more feedback. Individual crowd users reached a final average concordance of 80-81% after seeing around 75 test set clips (Figure 7). Experts maintained a relatively constant concordance level with the reference standard regardless of the number of test cases seen. The subset of skilled crowd users (n=114) showed an intermediate effect, having a slightly higher concordance than the overall crowd at all times and a less dramatic concordance increase, but still reaching a maximum individual concordance of around 81%.

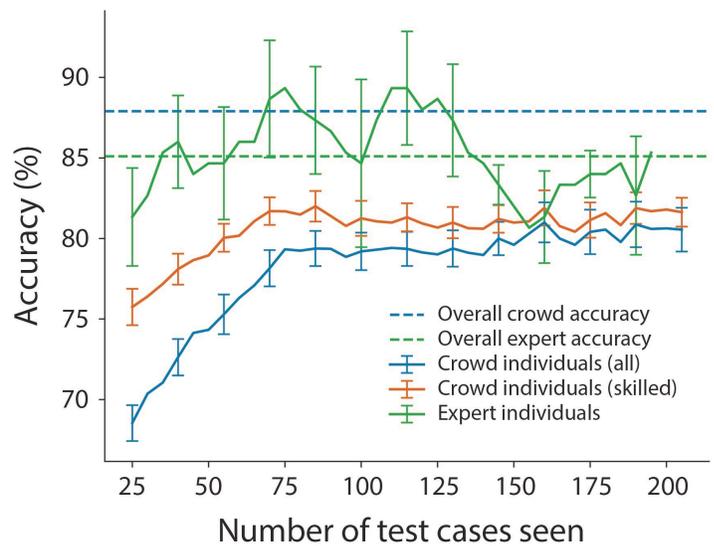

**Figure 7.** Learning curves for crowd and experts as measured by trailing accuracy relative to reference standard, plotted versus the number of test set cases seen. A window of 25 cases was used to compute trailing averages. Error bars represent standard error of the mean.

**LIMITATIONS**

Our dataset had an oversampling of clips with no lung pathology with more than 50% of clips in both the training and test datasets containing no B-lines. Since the crowd demonstrated higher concordance with reference standard than individual experts on clips with no B-lines, but performed worse than experts on classifying discrete B-lines, it is possible that the crowd may be less efficient at identifying subtle diagnostic findings, but we are not adequately seeing this trend due to dataset bias. While the similar or higher balanced multiclass accuracy of the crowd compared to individual experts supports the conclusion that crowd performance was on par with experts, future work will apply more balanced datasets.

While this labeling iOS application is available to any user, it is possible that users with medical background (medical students, etc.) are more likely to engage in this activity. We do have a general understanding of the proportion of our users with medical background, however the precise demographic breakdown of prior medical experiences was poorly defined. It is possible that our crowd is not representative of the general population and may be more consistent with a population of semi-skilled labelers. The generalizability of our findings across variable crowd populations with clearly defined experience levels will need to be explored further.

Our findings do show promise in streamlining labeling of lung POCUS data; however, this work may not be generalizable to other more complex medical data labeling tasks. Next steps are to apply this approach to similar questions in lung POCUS data such as segmenting B-lines, or evaluating for alternative findings beyond B-lines in lung POCUS. This will help us understand more about the generalizability of our labeling approach.

**DISCUSSION**

Our data suggest that gamified crowdsourcing can offer an efficient means of obtaining expert-quality labels for B-line classification on lung POCUS clips. Individual expert accuracies for classifying B-lines ranged considerably. This finding could have two possible explanations. First, given that medical imaging data interpretation is often complex, it may be that a proportion of clips have inherent ambiguity that even experts disagree on. This is consistent with previously published work that shows expert interrater agreement for identifying B-lines while substantial, is often imperfect.[39-42] Further, crowd inaccuracy and internal disagreement correlated with expert internal disagreement. Thus, it is possible that clips demonstrating crowd "inaccuracy" may actually be clinically equivocal rather than reflecting poor skill by the crowd. Inherent ambiguity is likely a theme that extends across medical imaging data beyond lung POCUS and may represent a challenge for imaging database labeling overall.

Variability in expert concordance could also be attributable to variable expert baseline skill. Consistent with existing literature which uses medical experts to label POCUS images, our experts all had either fellowship-level training or advanced certification in interpreting lung ultrasound as well as years of clinical experience.[28-33] Ground truth labeling for training POCUS-based ML models is commonly derived from a small handful of experts (typically 1-5 individuals). Thus, this work combining opinions from 6 experts (or 5 experts in the case of "leave-one-out consensus") to form our reference standard is consistent with accepted practices. Currently there is no established method for defining ground truth in B-line identification beyond expert opinion. Given the recent widespread adoption of lung POCUS globally and the recognized utility of B-lines as a clinical disease marker, our work highlights the critical need for clarifying how ground truth interpretation of lung POCUS and POCUS overall is defined.

Individual crowd users improved by their opinions converging toward the reference standard over the course of contest participation whereas expert concordance with the reference standard remained high throughout. This is expected as experts should not gain significant benefit beyond their baseline skill level from additional cases and thus are not expected to demonstrate a learning curve. This suggests that non-expert users can dynamically adapt to task-specific instructions or features based on real-time feedback. Even with tutorial- and feedback-based learning, individual crowd users generally did not improve to the concordance level of the average individual expert, but the combination of crowd users was able to exceed individual expert concordance. This illustrates the "wisdom of the crowd" effect of gaining accuracy from incorporating multiple opinions from users who are individually less skilled than experts to reach an accuracy level equal to that of experts.[43]

While commercial crowdsourcing platforms which allow users to perform discrete, repetitive tasks such as data labeling exist, our approach increased labeling accuracy and streamlined worker skill assessment via gamification. Gamified contests are structured to ensure crowd users put forth maximal effort, and offer users the opportunity for skill enhancement and learning via immediate feedback. Contest mechanics dynamically estimate user skill levels even as they fluctuate with learning from a subset of expert-annotated clips. The combination of continual user learning and selectively filtering for the most skilled users' opinions enhances labeling accuracy. While this approach does initially require a small number of expert-quality labels for crowd training, this input is a significant improvement in expert effort over what is commonly required for dataset labeling today.

The ultimate goal of our work is to identify strategies for scalable and accurate medical data labeling. When crowd opinions differed from each other, the split in crowd votes was a useful predictor of the expert-consensus label. Taken together with possible inherent ambiguity in some of the clips, this highlights a possible triage approach to dataset labeling using gamified crowdsourcing. Crowd opinion for clips with a high degree of crowd agreement would be accepted as truth, and expert review would only be necessary for cases where crowd agreement drops below a certain threshold. This approach could significantly decrease the proportion of clips requiring expert review and optimize both time and cost associated with current expert-based dataset labeling approaches.

Here we demonstrate that gamified crowdsourcing with strategic inbuilt quality control measures can produce B-line classification labels that match expert consensus better than individual experts themselves. These methods illustrate general strategies for improving the reliability of crowdsourced opinions on the task of classifying B-lines on lung POCUS clips. Using innovative and scalable approaches to generate high-quality labeled image databases could contribute to streamlining ML model development which could help either standardize or automate lung POCUS interpretation in the future.